\documentclass[12pt]{article}

\usepackage{amssymb}
\usepackage{amsmath}
\usepackage{graphicx}
\usepackage{tabls}
\usepackage{cite}

\newcommand{\dis}{\displaystyle}
\newcommand{\n}{\nonumber}
\newcommand{\N}{\normalsize}

\newcommand{\e}{{\rm e}}
\newcommand{\sgn}{{\rm sgn}}
\newcommand{\B}{{\cal B}}
\newcommand{\Q}{{\cal Q}}
\newcommand{\Z}{{\mathbb Z}}
\newcommand{\cO}{{\cal O}}
\newcommand{\cH}{{\cal H}}
\newcommand{\tg}{\tilde{g}}
\newcommand{\tf}{\tilde{f}}
\newcommand{\tR}{\tilde{R}}
\newcommand{\bx}{\bar{x}}

\newcommand{\Sin}{\,{\rm Sin(h)}}
\newcommand{\Cot}{\,{\rm Cot(h)}}

\newcommand{\vev}[1]{\left\langle #1 \right\rangle}

\def\coloneqq{\mathrel{\mathop:}=}

\makeatletter
\def\firstpage{\hfill KUNS-2573}
\def\ps@titlepage{
   \@oddhead{\hfil\firstpage\hfil}
   \def\@oddfoot{\hfil}}

\renewcommand{\thefootnote}{\,\fnsymbol{footnote}\,}
\makeatother

\title{\bf de Sitter Thin Brane Model\\[0.5em]}
\author{Masato Nishi\footnote{Email:\ masato@gauge.scphys.kyoto-u.ac.jp}\\[1em] \N \it Department of Physics, Kyoto University, Kyoto 606-8502, Japan}
\date{\vspace{0em}\N (August 26, 2015)}

\begin{document}
\maketitle

\baselineskip=16.7pt plus 0.2pt minus 0.1pt


\setcounter{page}{0}
\begin{abstract}
We discuss the large mass hierarchy problem in a braneworld model which represents our acceleratively expanding universe.
The Randall-Sundrum (RS) model with warped one extra dimension added to flat 4-dimensional space-time cannot describe our expanding universe.
Here, we study instead the de Sitter thin brane model.
This is described by the same action as that for the RS model, but the 4-dimensional space-time on the branes is $\rm dS_4$. 
We study the model for both the cases of positive 5-dimensional cosmological constant $\Lambda_5$ and negative one.
In the positive $\Lambda_5$ case, the 4-dimensional large hierarchy necessitates a 5-dimensional large hierarchy, and we cannot get a natural explanation.
On the other hand, in the negative $\Lambda_5$ case, the large hierarchy is naturally realized in the 5-dimensional theory in the same manner as in the RS model.
Moreover, another large hierarchy between the Hubble parameter and the Planck scale is realized by the $\cO(10^2)$ hierarchy of the 5-dimensional quantities.
Finally, we find that the lightest mass of the massive Kaluza-Klein modes and the intervals of the mass spectrum are of order $10^2\,\rm GeV$, which are the same as in the RS case and do not depend on the value of the Hubble parameter.

\end{abstract}
\thispagestyle{titlepage}
\newpage


\renewcommand{\thefootnote}{\,\arabic{footnote}\,}

\setcounter{section}{0}
\setcounter{subsection}{0}
\setcounter{figure}{0}

\section{Introduction}

In the last 15 years, it has been believed that braneworld models may give a solution to the problem of the extremely large hierarchy between the Planck scale, $10^{19}\,\rm GeV$, and the weak scale, $10^2\,\rm GeV$.
In 1998, Arkani-Hamed, Dimopoulos and Dvali first tried resolving the hierarchy problem by using a braneworld model\cite{ArkaniHamed:1998rs,Antoniadis:1998ig}.
Their explanation is as the follows:
First, they assume a flat $(4+d)$-dimensional space-time, in which the extra $d$-dimensions are compactified by a common radius $L$.
The relation between the $(4+d)$-dimensional Planck mass $M_{{\rm pl}(4+d)}$ and the 4-dimensional one $M_{\rm pl}$ is given by
\begin{align}
M^{2}_{\rm pl}\sim M^{d+2}_{{\rm pl}(4+d)}L^{d}. \label{eq:Boo}
\end{align}
Second, they assume that the fundamental scale is only the TeV scale, $1\,\rm TeV$.
Therefore, $M_{{\rm pl}(4+d)}$ should be of order $1\,\rm TeV$.
This means that a too large $M_{\rm pl}$ is specific to the 4-dimensional theory, and the large hierarchy does not exist in higher dimensions.
For example, if we take the number of the extra dimensions $d$ as $2$, Eq.\ \eqref{eq:Boo} implies that the radius $L$ is approximately $1\,\rm mm$.
This does not contradict experiments, since the Newton's law has been verified only at distances larger than $1\,\rm cm$.
However, note that the energy scale of $1/L$ is much smaller than the weak scale.\footnote
{If we take $M_{{\rm pl}(4+d)}\sim 1/L \sim 1\,\rm TeV$ to avoid a hierarchy in the $(4+d)$-dimensions, we have $M_{\rm pl}\sim 1\,\rm TeV$ independent of the number of the extra dimensions $d$.}
In other words, the hierarchy between the Planck scale and the weak scale is just replaced by the one between the weak scale and the radius $L$.

In keeping with this problem, Randall and Sundrum proposed a new braneworld model (the RS model)\cite{Randall:1999ee,Randall:1999vf}.\footnote
{In this paper, ``the RS model'' precisely means the RS1 model\cite{Randall:1999ee} .
Here, we do not refer to the RS2 model\cite{Randall:1999vf}.}
First, they assume a 5-dimensional space-time with negative 5-dimensional cosmological constant $\Lambda_5$ and warped extra dimension $y$.
Moreover, the extra dimension is compactified by $S^1/\Z_2$ with radius $L$.
Under these assumptions, the metric is given by
\begin{align}
ds^2=\e^{-2k|y|}\eta_{\mu\nu}dx^\mu dx^\nu+dy^2.
\end{align}
Hence, two branes are {\it naturally} introduced at the fixed points, $y=0,L$.
Second, they assume the energy scales (the vacuum expectation values of the Higgs boson) on the two branes to be $10^{19}\,\rm GeV$ (the Planck scale) and $10^2\,\rm GeV$ (the weak scale), respectively.
This means that the hierarchy does not exist on the former brane (the Planck brane), since its energy scale is equal to the 4-dimensional Planck mass $M_{\rm pl}$.
We live on the latter brane (the TeV brane), where the energy scale is equal to the weak scale.
Then, the large hierarchy on the TeV brane is realized by the $\cO(10)$ hierarchy, $kL\sim 39$, between the 5-dimensional quantities $k$ and $1/L$.
Moreover, we find that the 5-dimensional Planck mass $M_{{\rm pl}(5)}$ is of the same order as $k$ and $1/L$.
In this way, the 4-dimensional hierarchy is {\it naturally} explained in the 5-dimensional theory without any unnaturally large hierarchies among the 5-dimensional quantities.

However, the RS model does not represent our acceleratively expanding universe, since the 4-dimensional space-time on the branes is assumed to be flat.
Thus, it is necessary to study the models in which the 4-dimensional space-time on our brane is $\rm dS_4$.
Of course, there have been some papers studying such models, for example, \cite{Kaloper:1999sm,Parikh:2000fn,Neupane:2009ws,Neupane:2010ey}.
In the later two papers\cite{Neupane:2009ws,Neupane:2010ey}, the author assumes one warped and uncompactified extra dimension $y$ with negative $\Lambda_5$.
In addition, he introduces only a single brane with $\rm dS_4$ space-time.
Then, the 5-dimensional metric of the model is given by
\begin{align}
ds^2=\frac{H^2}{k^2}\sinh^2(k(|y|+\xi))\left\{-dt^2+\e^{2Ht}\eta_{ij}dx^idx^j\right\}+dy^2, \label{eq:Mario}
\end{align}
where $H$ is the Hubble parameter and $\xi$ is an arbitrary constant.
In this model, the relation between $M_{{\rm pl}(5)}$ and $M_{\rm pl}$ is given by
\begin{align}
M^2_{\rm pl}=M^3_{{\rm pl}(5)}\frac{H^2}{k^2}\int^{+\infty}_{-\infty} dy\,\sinh^2(k(|y|+\xi)). \label{eq:Luigi}
\end{align}
However, this is divergent at $y=\pm\infty$, and therefore, we cannot discuss the hierarchy problem in this model.


\subsection{Motivation for our work}

The above divergence problem of \eqref{eq:Luigi} is from the infinitely large integration range, namely, the uncompactified extra dimension.
If we compactify the extra dimension by $S^1/\Z_2$ similarly to the RS model, two branes are naturally introduced at the fixed points.
Then, the integration in \eqref{eq:Luigi} becomes finite, and we can discuss the hierarchy problem.
In other words, for making it possible to discuss the hierarchy problem, we need to introduce (at least) two branes.
In addition, since we have no apriori reason to restrict ourselves to the $\Lambda_5<0$ case, we will discuss the hierarchy problem for both the cases $\Lambda_5>0$ and $\Lambda_5<0$.
We call this model where the space-time on the branes is $\rm dS_4$ ``the de Sitter {\it thin} brane model'' for brevity.
Here, we have used the word ``thin'' to distinguish our model from {\it thick} brane models with smooth warp factors\cite{Dzhunushaliev:2009va,HerreraAguilar:2010kt,Barbosa-Cendejas:2013cxa}.

We should also examine the Kaluza-Klein modes for verifying the existence of the massless mode corresponding to the graviton, and non-existence of the light massive modes which could affect the Newton's law.
For this purpose, we need the exact expression of the wave function of the Kaluza-Klein modes, and this was constructed, for example, in\cite{Parikh:2000fn,Mannheim:2006qr}.
In the RS model, both the lightest mass of the massive Kaluza-Klein modes and the intervals of the mass spectrum are of order $10^2\,\rm GeV$, which suggests the possibility of the massive Kaluza-Klein particles being observed.
It is interesting to study the observability of the Kaluza-Klein modes in our model.

Actually, in \cite{Kaloper:1999sm,Parikh:2000fn}, the authors compactify the extra dimension and introduce two branes at the fixed points in the de Sitter thin brane model.
They discuss the hierarchy problem and the Kaluza-Klein modes in the model.
However, the Hubble parameter was not accurately determined at that time.
The aim of this paper is to give a complete analysis of the hierarchy problem and the Kaluza-Klein modes in the de Sitter thin brane model by using the observed value of the Hubble parameter.
Since our real universe has experienced much more complicated time evolution, namely,
big-bang $\rightarrow$ 
inflation $\rightarrow$ 
reheating $\rightarrow$ 
deceleration $\rightarrow$ 
accelerative expansion (current),
we should introduce the time-dependent Hubble parameter.
However, in this paper, for simplicity, we focus only on the current universe with a constant Hubble parameter.\footnote
{In \S4, we will comment on possible ways to make the Hubble parameter time-dependent in the de Sitter thin brane model.}

The parameter $\xi$ appearing in the metric \eqref{eq:Mario} plays an important role in our model.
Before compactifying the $y$-direction to $S^1/\Z_2$, $\xi$ is merely the freedom of translation in the $y$-direction.
However, after the compactification, if we put $\xi$ equal to zero, we will find that the brane tension is divergent and the model turns out to be {\it sick}.
Thus, for keeping the model {\it sound}, we must take non-zero $\xi$.
Moreover, we will also find that the parameter $\xi$ is important in naturally explaining the large hierarchy in our model, since, after the compactification, the integration \eqref{eq:Luigi} is finite and depends on $\xi$.


\subsection{Outline of the results}

We will find the following three important facts for our de Sitter thin brane model:
Recall that two branes are naturally introduced in the present model similarly to the RS model.
First, in the case with $\Lambda_5<0$, if we choose the energy scales on the two branes as $10^{19}\,\rm GeV$ and $10^2\,\rm GeV$, respectively, the 5-dimensional quantities can be almost of the same order; namely, the 4-dimensional large hierarchy is naturally realized in the 5-dimensional theory similarly to the RS model.
However, for $\Lambda_5>0$, we find that the large hierarchy in 4-dimensions necessarily implies a large hierarchy in 5-dimensions.
Thus, we conclude that we must choose $\Lambda_5<0$ to explain the hierarchy naturally.

Second, for $\Lambda_5<0$, the lightest non-zero mass of the Kaluza-Klein modes and the intervals of the mass spectrum are both of order $10^2\,\rm GeV$, which are insensitive to the value of the Hubble parameter.
Therefore, the lightest mass and the intervals in our model are the same as those in the RS model.

Finally, we find the importance of the parameter $\xi$ as we have already mentioned above.
Such a parameter certainly exists in the RS model.
However, it only effects a constant multiplication to the warp factor, which can be offset by a rescaling of $x^\mu$.
On the other hand, in our model, $\xi$ has a physical meaning.
Especially, $\xi$ is related to the Hubble parameter $H$.
Another 4-dimensional large hierarchy between $H$ and the Planck scale is realized by the $\cO(10^2)$ hierarchy, $k\xi\sim 102$, between the 5-dimensional quantities $k$ and $1/\xi$.

From the above results, we conclude that our thin brane model with $\Lambda_5<0$ can represent our acceleratively expanding universe, and at the same time naturally explain the large hierarchy.

The organization of this paper is as follows:
In {\S}2, we quickly review the RS model, with emphasis on the explanation of the hierarchy and the analysis of the Kaluza-Klein modes.
In {\S}3, we introduce the de Sitter thin brane model and study various properties of it:
explanation of the hierarchy problem, analysis of the graviton modes, and consideration of the important parameter $\xi$.
Finally, in {\S}4, we discuss the possibility of making the Hubble parameter time-dependent in our model.


\setcounter{section}{1}
\setcounter{subsection}{0}
\setcounter{figure}{0}

\section{Quick review of the RS model}

In this section, we briefly summarize  the RS model for use in later sections.


\subsection{Setup and the classical solution}

Let us consider the 5-dimensional space-time $x^M=(x^\mu,y)$ described by the following bulk action with negative cosmological constant $\Lambda_5$:
\begin{align}
S=M^3_{{\rm pl}(5)}\int d^5x \sqrt{-g}\,\left(R-2\Lambda_5\right), \label{eq:Bulbasaur}
\end{align}
where $M_{{\rm pl}(5)}$ is 5-dimensional Planck mass.
Under the metric assumption,
\begin{align}
ds^2 
=g_{MN}(y)dx^{M}dx^{N}
= \e^{-2A(y)}\eta_{\mu\nu}dx^{\mu} dx^{\nu}+dy^2, \label{eq:Ivysaur}
\end{align}
the solution to the Einstein equation with the condition that the extra dimension $y$ is compactified by $S^1/\Z_2$ with radius $L$ is given by
\begin{align}
A(y)=k|y|\quad (y\sim y+2L), \label{eq:Venusaur}
\end{align}
where the constant $k$ is defined by\footnote
{The replacement $k\rightarrow-k$ is essentially equivalent to $y\rightarrow L-y$.
Hence, we restrict $k$ to positive. \label{fn:Chansey}}
\begin{align}
k=\sqrt{-\frac{\Lambda_5}{6}}>0. \label{eq:Charmander}
\end{align}
The absolute value in \eqref{eq:Venusaur} is due to the compactification, and it causes the delta functions, $\delta(y)$ and $\delta(y-L)$, in $A''(y)$.
Because of these extra delta function terms, the Einstein equation is in fact {\em not} satisfied at $y=0,L$.
To canceling the extra terms, we introduce the following two rigid brane actions located at $y=0,L$:
\begin{align}
S_a
=-\lambda_a \int d^4x \sqrt{-g_a}
=-\lambda_a \int d^5x \sqrt{-g_a}\,\delta(y-y_a)\quad (a=1,2).
\end{align}
Here, $y_a$ is the brane position,
\begin{align}
y_1=0,\quad y_2=L, \n
\end{align}
and $g_a$ is the induced metric on the brane,
\begin{align}
{(g_a)}_{\mu\nu}=\left. \dis\frac{\partial x^M}{\partial x^{\mu}}\frac{\partial x^N}{\partial x^{\nu}}g_{MN}\right|_{y=y_a}=\e^{-2k|y_a|}\eta_{\mu\nu}. \label{eq:Charmeleon}
\end{align}
From the requirement that the Einstein equation of the whole system $S+S_1+S_2$ holds at $y=0,L$, the tension $\lambda_a$ is determined as
\begin{align}
\lambda_1=-\lambda_2=\frac{6k}{\kappa^2_5}=-\frac{\Lambda_5}{k}.
\end{align}
In this way, the compactification to $S^1/\Z_2$ {\em naturally} introduces the branes.


\subsection{Exponential hierarchy}

In this subsection, we describe how the hierarchy is naturally explained in the RS model.
We introduce the Higgs field $\cH$ with symmetry breaking {\em Mexican hat like} potential on each of the two branes located at $y=0,L$:
\begin{align}
S^{\rm \cH}_a=\int d^4x \sqrt{-g_a}\left[(g_a)^{\mu\nu}D_{\mu}\cH^{\dag}D_{\nu}\cH-\left(\cH^{\dag}\cH-v^2\right)^2\right]. \label{eq:Wigglytuff}
\end{align}
Since the coefficient of the kinetic term $\e^{-2k|y_a|}$ (see \eqref{eq:Charmeleon}) is not equal to $1$ for the second brane, we redefine $\cH$ to normalize it.
Then, the vacuum expectation value $v_a$ of the redefined field $\tilde{\cH}$ on the brane $a$ is expressed as
\begin{align}
v_a=\tilde{\vev{\cH}}_a=v\e^{-k|y_a|}, \label{eq:Fearow}
\end{align}
and their ratio is given by
\begin{align}
\frac{v_1}{v_2}=\e^{kL}. \label{eq:Charizard}
\end{align}
Note that $v_a$ is regarded as the energy scale on the brane $a$.
Now, we assume that the energy scale $v_1$ on the first brane is of the order of the Planck scale, $v_1=M_{\rm pl}\sim 10^{19}\,{\rm GeV}$.
And we assume also that our universe is on the second brane and therefore that $v_2=M_{\rm w}\sim 10^{2}\,{\rm GeV}$ (the weak scale).
Then, \eqref{eq:Charizard} implies that
\begin{align}
kL\simeq 39=\cO(10), \label{eq:Mew}
\end{align}
which is neither too large nor too small.
From now on, we call the brane at $y=0$ ($y=L$) ``the Planck brane'' (``the TeV brane'').

To understand how the 4-dimensional hierarchy is naturally realized in the 5-dimensional theory, we need to drive the relation between the 4-dimensional Planck mass $M_{\rm pl}$ and the 5-dimensional one $M_{{\rm pl}(5)}$.
For this purpose, we add to the metric \eqref{eq:Ivysaur} the following perturbation which does not depend on $y$:\footnote
{Such a restriction, of course, lacks the generality.
However, as we see in {\S}2.3, the zero-mode corresponding to the massless graviton does not depend on $y$.}
\begin{align}
g_{MN}(y)\ \rightarrow\ \tg_{MN}(x,y)
=g_{MN}(y)+h_{MN}(x)
= \left[
\begin{array}{cc}
\e^{-2A(y)}(\eta_{\mu\nu}+h_{\mu\nu}(x))\ \  &0 \\[0.5em]
0\ \ &1
\end{array}
\right], \label{eq:Squirtle}
\end{align}
where we have chosen the RS gauge with
\begin{align}
h_\mu^\mu=\partial^\nu h_{\mu\nu}=h_{M5}=h_{5M}=0. \label{eq:Eevee}
\end{align}
Then the 5-dimensional action \eqref{eq:Bulbasaur} (without the cosmological constant term) is reduced to
\begin{align}
M^3_{{\rm pl}(5)}\int d^5x \sqrt{-\tg}\,\tR(x,y)
=\frac{M^3_{{\rm pl}(5)}}{k}(1-\e^{-2kL})\int d^4x\sqrt{-\tf}\,\tR^{\rm 4D}(x), \label{eq:Wartortle}
\end{align}
where the 4-dimensional metric $\tf_{\mu\nu}$ is given by
\begin{align}
\tf_{\mu\nu}(x)=\eta_{\mu\nu}+h_{\mu\nu}(x),
\end{align}
and $\tR$ and $\tR^{\rm 4D}$ are made from $\tg_{MN}$ and $\tf_{\mu\nu}$, respectively.
From \eqref{eq:Wartortle}, we get the following relation between the 4-dimensional Planck mass and the 5-dimensional one:
\begin{align}
M^2_{\rm pl}=\frac{M^3_{{\rm pl}(5)}}{k}(1-\e^{-2kL})\simeq \frac{M^3_{{\rm pl}(5)}}{k}. \label{eq:Blastoise}
\end{align}
Note that we can neglect $\e^{-2kL}$ since we are taking $kL\simeq39$.
The relation \eqref{eq:Blastoise} is, for example, realized by taking
\begin{align}
M_{{\rm pl}(5)}\sim k\sim M_{\rm pl}\sim 10^{19}\,{\rm GeV}. \label{eq:Caterpie}
\end{align}
In this case, all the 5-dimensional quantities, $M_{{\rm pl}(5)}$, $k$, and $1/L$, are approximately equal to $M_{\rm pl}$, and this is a welcome result.
Thus, the 4-dimensional hierarchy $M_{\rm pl}/v_2\sim 10^{17}$ is realized without introducing unnatural 5-dimensional hierarchies.

Then, is the Planck scale the unique 5-dimensional fundamental scale?
The answer is {\em no}.
For example, let us adopt a new coordinate $\bx^{\mu}$ related to the original $x^{\mu}$ by the scale transformation
\begin{align}
\bx^{\mu}= \e^{-kL}x^{\mu}. \label{eq:Metapod}
\end{align}
The warp factor for the coordinate $\bx^{\mu}$,
\begin{align}
\e^{2k(L-|y|)}, \label{eq:Dragonite}
\end{align}
 is normalized at $y=L$ (the TeV brane position).
Since $d^4x$ and $\tR^{\rm 4D}(x)$ are equal to $\e^{+4kL}d^4\bx$ and $\e^{-2kL}\tR^{\rm 4D}(\bx)$, respectively, the 5-dimensional action \eqref{eq:Wartortle} is now given by
\begin{align}
\frac{M^3_{{\rm pl}(5)}}{k}(\e^{+2kL}-1)\int d^4\bx\sqrt{-\tf}\,\tR^{\rm 4D}(\bx).
\end{align}
This implies that the relation \eqref{eq:Blastoise} is modified to
\begin{align}
M^2_{\rm pl}=\frac{M^3_{{\rm pl}(5)}}{k}(\e^{+2kL}-1)\simeq \frac{M^3_{{\rm pl}(5)}}{k}\e^{+2kL}. \label{eq:Butterfree}
\end{align}
In this case, $M_{\rm pl}\sim10^{19}\,\rm GeV$ is realized by taking
\begin{align}
M_{{\rm pl}(5)}\sim k\sim M_{\rm w}\sim 10^{2}\,{\rm GeV}, \label{eq:Weedle}
\end{align}
in which all the 5-dimensional quantities are approximately of the weak scale.
In this way, we can arbitrarily change the energy scale of all the 5-dimensional quantities through the transformation.
In the above two examples with the coordinates $x^{\mu}$ and $\bx^{\mu}$, all the 5-dimensional quantities are of the scale of the brane at which the warp factor is normalized.


\subsection{Graviton modes}

Since the extra dimension $y$ is compactified, there appear the towers of the Kaluza-Klein modes.
We must verify the existence of the massless graviton corresponding to the zero-mode.
We must also verify the non-existence of light massive modes which could affect the Newton's law. 
Let us consider the perturbation \eqref{eq:Squirtle} with $h_{MN}$ now depending on $y$ as well as on $x^{\mu}$.
Under the assumption that $h_{\mu\nu}$ can be expanded as
\begin{align}
h_{\mu\nu}(x,y)=\e^{\frac{3}{2}A(y)}\sum^{\infty}_{n=0}\phi^{(n)}_{\mu\nu}(x)\psi_n(y), \label{eq:Kakuna}
\end{align}
with $\phi^{(n)}_{\mu\nu}(x)$ being the 4-dimensional field with mass $m_n$, the modes $\psi_n(y)$ excluding the points $|y|=0,L$ are given by
\begin{subequations}\label{eq:Beedrill}
\begin{align}
\psi_0(y)&=a_0\e^{-\frac{3}{2}A(y)}, \label{eq:Pidgey} \\[0.5em]
\psi_n(y)&=\sqrt{\frac{\e^{k|y|}}{k}}\left\{a_nJ_2\left(m_n\e^{k|y|}/k\right)+b_nY_2\left(m_ne^{k|y|}/k\right)\right\} \quad (n\geq1). \label{eq:Pidgeotto}
\end{align}
\end{subequations}
Here, $J_\alpha$ and $Y_\alpha$ are the Bessel functions of the first and second kind, respectively, and $a_n$ and $b_n$ are constants.
In particular, $\psi_0(y)$ is the zero-mode with $m_0=0$.
Note that the $y$ dependences of the $n=0$ term in \eqref{eq:Kakuna} cancel.

The 4-dimensional mass $m_n$ and the ratio $a_n/b_n$ are determined by the boundary conditions at the brane positions.
These conditions are derived by integrating the differential equation for $\psi_n(y)$ in infinitesimal small regions containing the brane positions.
We find that $m_n$ for the massive modes are determined by the following equation:
\begin{align}
J_1(m_n/k)Y_1(m_n\e^{kL}/k)-J_1(m_n\e^{kL}/k)Y_1(m_n/k)=0. \label{eq:Rattata}
\end{align}
For a very small $m_n$ with $m_n/k\ll 1/e$,\footnote
{Here, the warp factor is normalized at the Planck brane, and we have $k\sim 10^{19}\,\rm GeV$.}
the first term of \eqref{eq:Rattata} can be neglected, since we have\begin{align}
J_1(m_n/k)\sim \frac{m_n}{k}\ll 1,\qquad
|Y_1(m_n/k)|\sim \left|\frac{m_n}{k}\log\left(\frac{m_n}{k}\right)\right|\gg \frac{m_n}{k}.
\end{align}
Therefore, $m_n$ is determined as
\begin{align}
J_1(m_n\e^{kL}/k)\simeq 0\ \Rightarrow\ m_n\simeq k\e^{-kL}j_n, \label{eq:Raticate}
\end{align}
where $j_n$ is the $n$-th zero of $J_1$.
Since the intervals of adjacent zeroes of $J_1$ are approximately equal to $\pi$, the mass difference $\Delta m_n$ is given by
\begin{align}
\Delta m_n = m_{n+1}-m_n\sim \pi\times 10^2\,{\rm GeV}. \label{eq:Spearow}
\end{align}
This result implies that the Newton's law remains unmodified for a scale larger than $10^{-18}\,\rm m$.
However, the first massive Kaluza-Klein particle could be observed in the near future.


\setcounter{section}{2}
\setcounter{subsection}{0}
\setcounter{figure}{0}

\section{de Sitter thin brane model}

The RS model assumes that the 4-dimensional space-time on the branes is static.
However, we know that our real universe is acceleratively expanding.
In this section, we construct a 5-dimensional braneworld model where the 4-dimensional space-time on the branes is $\rm dS_4$ describing our expanding universe.
The warp factor in our model has cusp singularities at the brane positions as in the RS model.
Therefore, we call our model ``the de Sitter {\it thin} brane model'' in contrast to ``the de Sitter {\it thick} brane model'' where the warp factor is smooth and has non-singularities\cite{Dzhunushaliev:2009va,HerreraAguilar:2010kt,Barbosa-Cendejas:2013cxa}.
As mentioned in {\S}1.1, we consider for simplicity only our current universe with a constant Hubble parameter.


\subsection{Setup and the classical solution}

Let us consider the 5-dimensional action \eqref{eq:Bulbasaur} and the following metric:
\begin{align}
ds^2=g_{MN}(t,y)dx^Mdx^N
&=\e^{2A(y)}f_{\mu\nu}(t)dx^\mu dx^\nu+dy^2 \n \\
&= \e^{2A(y)}\left\{-dt^2+a^2(t)\eta_{ij}dx^idx^j\right\}+dy^2. \label{eq:Victini}
\end{align}
Here, we consider both the cases of $\Lambda_5>0$ and $\Lambda_5<0$.
The main difference from the RS model of {\S}2 is that the 4-dimensional space-time is not static, but is the FLRW metric with flat space, which is the simplest metric describing our acceleratively expanding universe.
Under these assumptions, the Einstein tensor $G_{MN}$ is expressed as follows:
\begin{align}
G_{MN}&=3\left[A''+2A'^2-\left(\frac{\dot{a}}{a}\right)^2\e^{-2A}\right]g_{00}\delta^0_M\delta^0_N \n \\
&\hspace{3em}+\left[3(A''+2A'^2)-\left\{\frac{2\ddot{a}}{a}+\left(\frac{\dot{a}}{a}\right)^2\right\}\e^{-2A}\right]g_{ij}\delta^i_M\delta^j_N \n \\
&\hspace{3em}+3\left[2A'^2-\left\{\frac{\ddot{a}}{a}+\left(\frac{\dot{a}}{a}\right)^2\right\}\e^{-2A}\right]g_{55}\delta^5_M\delta^5_N, \label{eq:Servine}
\end{align}
where the overdots and the primes denote derivatives with respect to $t$ and $y$, respectively.
Thus, from the Einstein equation, we get the differential equations for $A(y)$ and $a(t)$:
\begin{align}
A''=-\frac{\ddot{a}}{a}\e^{-2A},\qquad 
A'^2=\frac{1}{2}\left[\frac{\ddot{a}}{a}+\left(\frac{\dot{a}}{a}\right)^2\right]\e^{-2A}-\frac{1}{6}\Lambda_5, \qquad
a\ddot{a}=\dot{a}^2. \label{eq:Serperior}
\end{align}
Plugging the solution of the last equation of \eqref{eq:Serperior},
\begin{align}
a(t)=\e^{Ht},
\end{align}
where $H$ is an arbitrary constant (the Hubble parameter), into the rest of the equations of \eqref{eq:Serperior}, we obtain
\begin{align}
A''=-H^2\e^{-2A},\qquad A'^2=H^2\e^{-2A}-\frac{1}{6}\Lambda_5. \label{eq:Gigalith}
\end{align}
Two differential equations in \eqref{eq:Gigalith} are not independent; the first is obtained by differentiating the second.
In any case, the general solution is given by
\begin{align}
A(y)=\log\left|\frac{H}{k}\Sin(k(y+\xi))\right|, \label{eq:Tepig}
\end{align}
where $\xi$ is an arbitrary constant, and $k$ in the present model is defined by\footnote
{Since \eqref{eq:Tepig} is invariant under the replacement $k\rightarrow-k$, we choose $k$ to be positive. \label{fn:Cryogonal}}
\begin{align}
k=\sqrt{\frac{|\Lambda_5|}{6}}>0. \label{eq:Pignite}
\end{align}
In \eqref{eq:Tepig}, we have introduced a new function ${\rm Sin(h)}(x)$ defined by
\begin{align}
\Sin(x)=
\begin{cases}
\sin x & (\Lambda_5>0) \\
\sinh x & (\Lambda_5<0)
\end{cases}
\end{align}
for treating the both cases $\Lambda_5\gtrless 0$ by a single equation.
Later we will also introduce $\rm Cot(h)$ defined similarly.

Now, to introduce two branes naturally, let us compactify the 5th-dimension by $S^1/\Z_2$ with radius $L$.
Then, the expressions of $A(y)$ and its derivatives are altered as follows:
\begin{align}
A(y)&=\log\left|\frac{H}{k}\Sin(k(|y|+\xi))\right|, \n \\
A'(y)&=k\,\sgn(y)\Cot(k(|y|+\xi)), \n \\
A''(y)&=-\frac{k^2}{\Sin^2(k(|y|+\xi))}+2k\Cot(k(|y|+\xi))\{\delta(y)-\delta(y-L)\}. \label{eq:Emboar}
\end{align}
Due to the delta function term in $A''(y)$, the Einstein equation is not satisfied at $y=0$ and $y=L$.
To compensate, we must introduce the following two brane actions $S_a\ \ (a=1,2)$:
\begin{align}
S_a=(-1)^a\frac{\Lambda_5}{k}\Cot(k(y_a+\xi)) \int d^4x \sqrt{-g_a}, \label{eq:Dewott}
\end{align}
where $y_{1,2}=0,L$ are the brane positions, and $g_a$ are the induced metrics on the branes:
\begin{align}
{(g_a)}_{\mu\nu}=\e^{2A(y_a)}f_{\mu\nu}. \label{eq:Samurott}
\end{align}


\subsection{Hierarchy problem}

Let us consider whether the above time-dependent model can explain the hierarchy problem.
We introduce the same Higgs action as \eqref{eq:Wigglytuff} on each of the two branes $a=1,2$.
For the present metric \eqref{eq:Victini}, it is given by
\begin{align}
S^{\rm \cH}_a= \int d^4x \sqrt{-f}\left[f^{\mu\nu}D_{\mu}\tilde{\cH}^{\dag}D_{\nu}\tilde{\cH}-\left\{\tilde{\cH}^{\dag}\tilde{\cH}-\left(\e^{A(y_a)}v\right)^2\right\}^2\right], \label{eq:Petilil}
\end{align}
where we have introduced $\tilde{\cH}=\e^{A(y_a)}\cH$ to normalize the kinetic term.
Thus, the vacuum expectation value $v_a$ of the field $\tilde{\cH}$ on the brane $a$ is given by
\begin{align}
v_a=\e^{A(y_a)}v, \label{eq:Lilligant}
\end{align}
which can be regarded as the energy scale on the brane $a$.
Using \eqref{eq:Lilligant}, the ratio $v_2/v_1$ is given by
\begin{align}
\frac{v_2}{v_1}=\left|\frac{\Sin(k(L+\xi))}{\Sin(k\xi)}\right|. \label{eq:Patrat}
\end{align}
Since the numerator and the denominator of \eqref{eq:Patrat} is exchanged under the replacement $\xi\rightarrow-(L+\xi)$, we can restrict ourselves to the case $v_1<v_2$ without loss of generality.
Now, let us consider the situation, $v_1=M_{\rm w}\sim 10^{2}\,{\rm GeV}$ and $v_2=M_{\rm pl}\sim 10^{19}\,{\rm GeV}$, which means that our universe is on the first brane and the 4-dimensional hierarchy dose not exist on the second brane.
Hereafter, we call the branes at $y=0$ and $y=L$ ``the TeV brane'' and ``the Planck brane'', respectively.
In the following, we consider whether the large hierarchy $v_2/v_1\sim10^{17}$ can be realized without introducing any unnatural hierarchies among the 5-dimensional quantities $k$, $L$ and $\xi$, for both the cases $\Lambda_5>0$ and $\Lambda_5<0$.\\


\noindent
\underline{$\Lambda_5>0$}\,:\\[0.5em]
In this case, our problem is how the condition
\begin{align}
\left|\frac{\sin(k(L+\xi))}{\sin(k\xi)}\right|\sim 10^{17}
\end{align}
can be naturally realized.
Examination of this condition for both the cases of $kL=\cO(1)$ and $kL\ll1$ (modulo integer multiples of $\pi$) leads to a single requirement
\begin{align}
\frac{L}{|\xi|}\sim 10^{17}, \label{eq:Lillipup}
\end{align}
namely, we need a fine tuning.\\


\noindent
\underline{$\Lambda_5<0$}\,:\\[0.5em]
From \eqref{eq:Patrat}, for not too small $k|\xi|$ ($k|\xi|\gg 10^{-17}$), we get\footnote
{In deriving \eqref{eq:Herdier}, we used the formula $\sinh^{-1}x=\log(x+\sqrt{x^2+1})$ to rewrite \eqref{eq:Patrat} as follows:
\begin{align}
k(L+\xi)&\sim\log\left(10^{17}\sinh(k|\xi|)+\sqrt{\{10^{17}\sinh(k|\xi|)\}^2+1}\right) \n \\
&\hspace{-1.7em}\overset{k|\xi|\gg 10^{-17}}{\simeq} \log(2\cdot 10^{17}\sinh(k|\xi|))
\simeq 39+\log\left|\e^{k\xi}-\e^{-k\xi}\right|. \n
\end{align}
Note that we have not made any restrictions on $kL$.}
\begin{align}
kL\sim 39+\log\left|1-\e^{-2k\xi}\right|. \label{eq:Herdier}
\end{align}


To verify whether the present model can explain the large hierarchy, we must calculate the relationship between the 5-dimensional Planck mass $M_{{\rm pl}(5)}$ and the 4-dimensional one $M_{\rm pl}$.
For this purpose, let us add a perturbation to the metric \eqref{eq:Victini} as follows:\footnote
{As we will see in the next subsection {\S}3.3, the zero-mode $h^{(0)}_{\mu\nu}$ corresponding to the massless graviton does not depend on $y$, similarly to the case of the RS model explained in {\S}2.3.}
\begin{align}
g_{MN}(t,y)\ \rightarrow\ \tg_{MN}(x,y)
&=g_{MN}(t,y)+h_{MN}(x) \n \\
&= \left[
\begin{array}{cc}
\e^{2A(y)}\{f_{\mu\nu}(t)+h_{\mu\nu}(x)\}\ \  &0 \\[0.5em]
0\ \ &1
\end{array}
\right], \label{eq:Liepard}
\end{align}
where we have taken the RS gauge \eqref{eq:Eevee}.
The Ricci scalar $\tilde{R}(x,y)$ made from the metric \eqref{eq:Liepard} is calculated as
\begin{align}
\tilde{R}(x,y)=\e^{-2A}\tilde{R}^{\rm 4D}(x)+(h\text{-independent term}),
\end{align}
where $\tilde{R}^{\rm 4D}(x)$ is the Ricci scalar made from $\tf_{\mu\nu}(x)\coloneqq f_{\mu\nu}(t)+h_{\mu\nu}(x)$.
Thus, the Ricci scalar part of the 5-dimensional action \eqref{eq:Bulbasaur} is given as follows:
\begin{align}
&M^3_{{\rm pl}(5)}\int d^5x\sqrt{-\tilde{g}}\,\tilde{R}(x,y)
=M^3_{{\rm pl}(5)}\int^{+L}_{-L}dy\,\e^{2A}\int d^4x\sqrt{-\tf(x)}\,\tilde{R}^{\rm 4D}(x) \n \\
&=\sgn(\Lambda_5)M^3_{{\rm pl}(5)}\frac{H^2}{k^3}\left\{kL+\frac{\Sin(2k\xi)-\Sin(2k(L+\xi))}{2}\right\}\int d^4x\sqrt{-\tf(x)}\,\tilde{R}^{\rm 4D}(x). \label{eq:Pansage}
\end{align}
From this, we get the following relationship between $M_{\rm pl}$ and $M_{{\rm pl}(5)}$:
\begin{align}
M^2_{\rm pl}=
\sgn(\Lambda_5)M^3_{{\rm pl}(5)}\frac{H^2}{k^3}\left\{kL+\frac{\Sin(2k\xi)-\Sin(2k(L+\xi))}{2}\right\}. \label{eq:Simisage}
\end{align}

In the following, we would like to take as the Hubble parameter $H$ the observed value.
However, the Hubble parameter $H$ can be varied by a rescaling of the 4-dimensions $x^\mu$.
When the warp factor is normalized at $y=0$ (the TeV brane position), the Hubble parameter $H$ should be equal to the observed value on the TeV brane $H_0\sim 10^{-42}\,\rm GeV$.
Therefore, we adopt a new 4-dimensional coordinate $\bx^\mu$ defined by
\begin{align}
\bx^\mu=\e^{A(0)}x^\mu=\frac{H}{k}\left|\Sin(k\xi)\right|\cdot x^\mu. \label{eq:Cobalion}
\end{align}
Then, the metric \eqref{eq:Victini} is modified to
\begin{align}
ds^2=\left(\frac{\Sin(k(|y|+\xi))}{\Sin(k\xi)}\right)^2\bar{f}_{\mu\nu}(\bar{t})d\bx^\mu d\bx^\nu+dy^2, \label{eq:Terrakion}
\end{align}
with
\begin{align}
\bar{f}_{\mu\nu}(\bar{t})
=\left[
\begin{array}{cc}
-1\ \  &0 \\[0.5em]
0\ \ &\e^{2H_0\bar{t}}\delta_{ij}
\end{array}
\right]
,\qquad H_0=\frac{k}{\left|\Sin(k\xi)\right|}. \label{eq:Virizion}
\end{align}
Accordingly, the relation between $M_{\rm pl}$ and $M_{{\rm pl}(5)}$ \eqref{eq:Simisage} is modified to
\begin{align}
M^2_{\rm pl}=
\sgn(\Lambda_5)M^3_{{\rm pl}(5)}\frac{H_0^2}{k^3}\left\{kL+\frac{\Sin(2k\xi)-\Sin(2k(L+\xi))}{2}\right\}, \label{eq:Tornadus}
\end{align}
namely, the rescaling is just equivalent to replacing $H$ with $H_0$.
Hereafter, when we use the observed value of the Hubble parameter, we rescale $x^\mu$ as \eqref{eq:Cobalion} and use $H_0$.

Now, we impose that $M_{{\rm pl}(5)}\sim k$, namely, a requirement of the absence of the 5-dimensional hierarchy.
Therefore, \eqref{eq:Tornadus} is rewritten as
\begin{align}
\sgn(\Lambda_5)\left\{kL+\frac{\Sin(2k\xi)-\Sin(2k(L+\xi))}{2}\right\}\sim 10^{122}, \label{eq:Woobat}
\end{align}
where we have used $M_{\rm pl}\sim 10^{19}\,\rm GeV$ and $H_0\sim 10^{-42}\,\rm GeV$.
We will examine \eqref{eq:Woobat} for the cases of $\Lambda_5>0$ and $\Lambda_5<0$.\\


\noindent
\underline{$\Lambda_5>0$}\,:\\[0.5em]
In this case, the second term inside the curly brackets of \eqref{eq:Woobat} is at most 1.
Therefore, we must take as $kL$ an extremely large value, $kL\sim 10^{122}$.\\


\noindent
\underline{$\Lambda_5<0$}\,:\\[0.5em]
In this case, from \eqref{eq:Herdier} and \eqref{eq:Woobat}, we obtain
\begin{align}
\frac{1}{2}\left\{\sinh\Bigr(2\left(k\xi+39+\log\left|1-\e^{-2k\xi}\right|\right)\Bigr)-\sinh(2k\xi)\right\}-39-\log\left|1-\e^{-2k\xi}\right|
\sim 10^{122}, \label{eq:Escavalier}
\end{align}
and solving this equation numerically, we get two solutions:
\begin{align}
(k\xi,kL)\sim (+102,39),\quad (-102,243). \label{eq:Pansear}
\end{align}
Both of these values are consistent with our assumption $k|\xi|\gg10^{-17}$, and at the same time show that $kL$ is neither too large nor too small.
As we will see in {\S}3.4.1, we must exclude the negative $\xi$ case, since the action of fluctuation diverges.
However, we will continue our argument without restricting ourselves to the positive $\xi$ case.
At this point, we conclude that the case $\Lambda_5<0$ is a candidate for solving the hierarchy problem.

In the above discussion, we did not mention the absolute values of the 5-dimensional quantities $M_{{\rm pl}(5)}$, $k$, $1/L$ and $1/\xi$.
However, we can fix the value of $k$ from the expression of $H_0$ \eqref{eq:Virizion} and \eqref{eq:Pansear} to obtain $k\sim 10^2\,\rm GeV$.
Consequently, all the absolute values of the 5-dimensional quantities are uniquely fixed to be almost of the same order $10^2\,\rm GeV$, which is equal to the weak scale $M_{\rm w}$ (the energy scale on the TeV brane).
In this respect, the de Sitter thin brane model is largely different from the RS model (see {\S}2.2);
the RS model lacks information that can uniquely fix the absolute values of the 5-dimensional quantities. 


\subsection{Graviton modes}

Next, let us study the Kaluza-Klein graviton modes in the present model.
In particular, we are interested in whether the massless graviton exists,\footnote
{Of course, we already know the existence of the massless graviton from the argument of {\S}3.2.}
and the effect of the massive modes on the Newton's law. 
Then, we consider the perturbed metric \eqref{eq:Liepard} with $h_{\mu\nu}$ now having the $y$-dependence as well as the $x^{\mu}$-dependence, $h_{\mu\nu}=h_{\mu\nu}(x,y)$.
Moreover, we assume that $h_{\mu\nu}$ can be expanded as
\begin{align}
h_{\mu\nu}(x,y)=\e^{-\frac{3}{2}A(y)}\sum^{\infty}_{n=0}\phi^{(n)}_{\mu\nu}(x)\psi_n(y), \label{eq:Simisear}
\end{align}
where $\phi^{(n)}_{\mu\nu}(x)$ is the 4-dimensional field with mass $m_n$.
The modes $\psi_n(y)$ have to satisfy the following differential equation \cite{Parikh:2000fn,Mannheim:2006qr}:
\begin{align}
&\psi''(y)+k\coth(k(|y|+\xi))\psi'(y) \n \\
&-\left\{\frac{15}{4}k^2+\frac{k^2M_n^2}{\sinh^2(k(|y|+\xi))}+3k\coth(k(|y|+\xi))\{\delta(y)-\delta(y-L)\}\right\}\psi(y)=0. \label{eq:Chandelure}
\end{align}
Here, we have defined $M_n$ as
\begin{align}
M_n= \sqrt{\frac{9}{4}-\frac{m_n^2}{H^2}}. \label{eq:Archeops}
\end{align}
The general solution to \eqref{eq:Chandelure} excluding $|y|=0,L$ is given by
\begin{align}
\psi_n(y)=a_nP^{M_n}_{3/2}\bigr(\cosh(k(|y|+\xi))\bigr)+b_nQ^{M_n}_{3/2}\bigr(\cosh(k(|y|+\xi))\bigr), \label{eq:Panpour}
\end{align}
where $P^\mu_\nu$ and $Q^\mu_\nu$ are the associated Legendre functions of the first and second kind, respectively.
Similarly to the case of the RS model of {\S}2.3, the 4-dimensional mass $m_n$ and the ratio $a_n/b_n$ are determined by the boundary conditions at the brane positions obtained by integrating \eqref{eq:Chandelure} in the infinitesimal small regions containing $y=0$ and $y=L$. 
These conditions are given by
\begin{align}
\psi'_n(y)\Bigr|_{y=+0}=\frac{3}{2}k\coth(k\xi)\psi_n(0),\qquad
\psi'_n(y)\Bigr|_{y=L-0}=\frac{3}{2}k\coth(k(L+\xi))\psi_n(L), \label{eq:Musharna}
\end{align}
namely,\footnote
{In deriving \eqref{eq:Yamask}, we have used the following recursion relation:
\begin{align}
(w^2-1)\frac{dP^\mu_\nu(w)}{dw}&=-(\nu+\mu)P^\mu_{\nu-1}(w)+\nu wP^\mu_\nu(w), \n 
\end{align}
and the same relation for $Q^\mu_\nu(w)$\cite{Math}.}
\begin{align}
a_n P^{M_n}_{1/2}\bigr(\cosh(k(y_a+\xi)\bigr)+b_n Q^{M_n}_{1/2}\bigr(\cosh(k(y_a+\xi))\bigr)=0. \label{eq:Yamask}
\end{align}
The masses $m_n$ are determined by the condition of the existence of non-trivial $(a_n,b_n)$:
\begin{align}
P^{M_n}_{1/2}\bigr(\cosh(k\xi)\bigr)Q^{M_n}_{1/2}\bigr(\cosh(k(L+\xi))\bigr)
-P^{M_n}_{1/2}\bigr(\cosh(k(L+\xi))\bigr)Q^{M_n}_{1/2}\bigr(\cosh(k\xi)\bigr)=0. \label{eq:Pidove}
\end{align}


\subsubsection{Zero-mode}

For the zero-mode with $m_0=0$ ($M_0=3/2$), we see that \eqref{eq:Pidove} is realized owing to the following relations:
\begin{align}
P^{3/2}_{1/2}\left(\cosh\eta\right)\propto
Q^{3/2}_{1/2}\left(\cosh\eta\right)\propto\frac{1}{\sinh^{\frac{3}{2}}|\eta|}. \label{eq:Tranquill}
\end{align}
Moreover, from \eqref{eq:Musharna} and the following relations,
\begin{subequations}\label{eq:Drilbur}
\begin{align}
&P^{3/2}_{3/2}\left(\cosh\eta\right)\propto\frac{3\cosh|\eta|-\cosh(3|\eta|)}{\sinh^{\frac{3}{2}}|\eta|}, \\[0.5em]
&Q^{3/2}_{3/2}\left(\cosh\eta\right)\propto\frac{3\cosh|\eta|-\cosh(3|\eta|)+4\sinh^3|\eta|}{\sinh^{\frac{3}{2}}|\eta|},
\end{align}
\end{subequations}
we see that the zero-mode $\psi_0(y)$ is given by
\begin{align}
\psi_0(y)\propto\sinh^{\frac{3}{2}}|k(|y|+\xi)|= \e^{\frac{3}{2}A(y)}.
\end{align}
Hence, from \eqref{eq:Simisear}, we find that the zero-mode part of $h_{\mu\nu}(x,y)$ does not depend on $y$:
\begin{align}
\e^{-\frac{3}{2}A(y)}\phi^{(0)}_{\mu\nu}(x)\psi_0(y)\propto\phi^{(0)}_{\mu\nu}(x).
\end{align}


\subsubsection{Massive modes}

\begin{figure}[t]
\begin{center}
\includegraphics[width=10cm,bb=0 0 400 280]{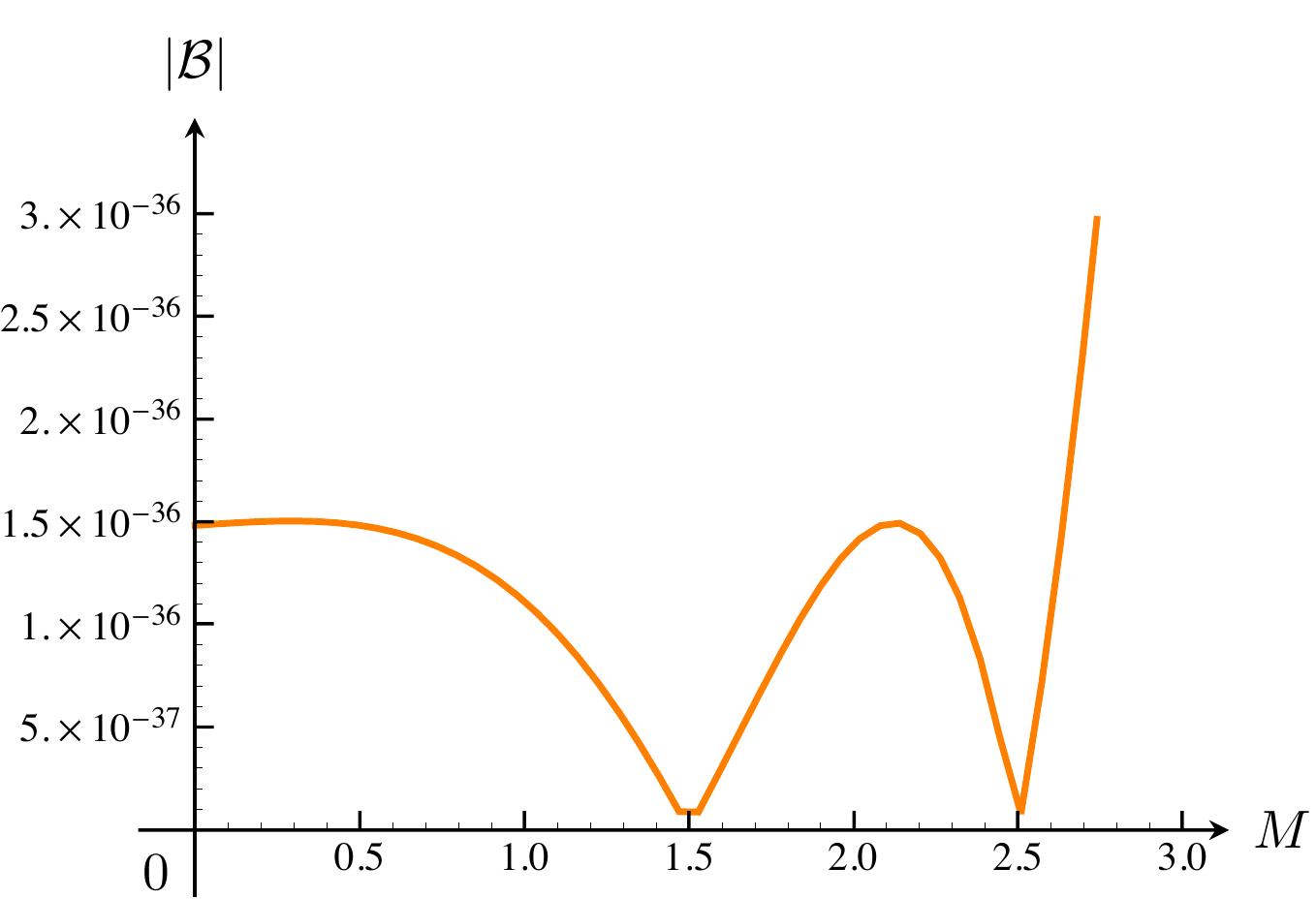}
\end{center}
\caption{The function \eqref{eq:Archen} for real $M$ ($m^2\leq\frac{9}{4}H^2$).
The zero point $M=3/2$ corresponds to the zero-mode $m=m_0=0$.
Though the curve is not properly displayed near $M=3/2$ and $5/2$ due to limitations of Mathematica, $|\B|$ exactly vanishes at those points.} \label{fig:Palpitoad}
\end{figure}

Now, let us consider the left-hand side of \eqref{eq:Pidove} with $M_n$ replaced with $M$, and denote it by $\B(M;k\xi,k(L+\xi))$.
We seek the zero points of $\B(M;k\xi,k(L+\xi))$ as a function of $M$.
Accordingly, we define another variable $m$ by $M=\sqrt{\frac{9}{4}-\frac{m^2}{H^2}}$.
For real $M$ ($m^2\leq \frac{9}{4}H^2$), we can numerically analyze
\begin{align}
\left|\B\Bigr(M;k\xi,k(L+\xi)=39+\log\left|\e^{k\xi}-\e^{-k\xi}\right|\Bigr)\right| \label{eq:Archen}
\end{align}
for the value of $k|\xi|$ given by \eqref{eq:Pansear}, $k|\xi|\sim 102$ ($kL$ is related to $k|\xi|$ by \eqref{eq:Herdier}), and the result is shown in Fig.\ \ref{fig:Palpitoad}.

\begin{figure}[t]
\begin{center}
\includegraphics[width=10cm,bb=0 0 400 280]{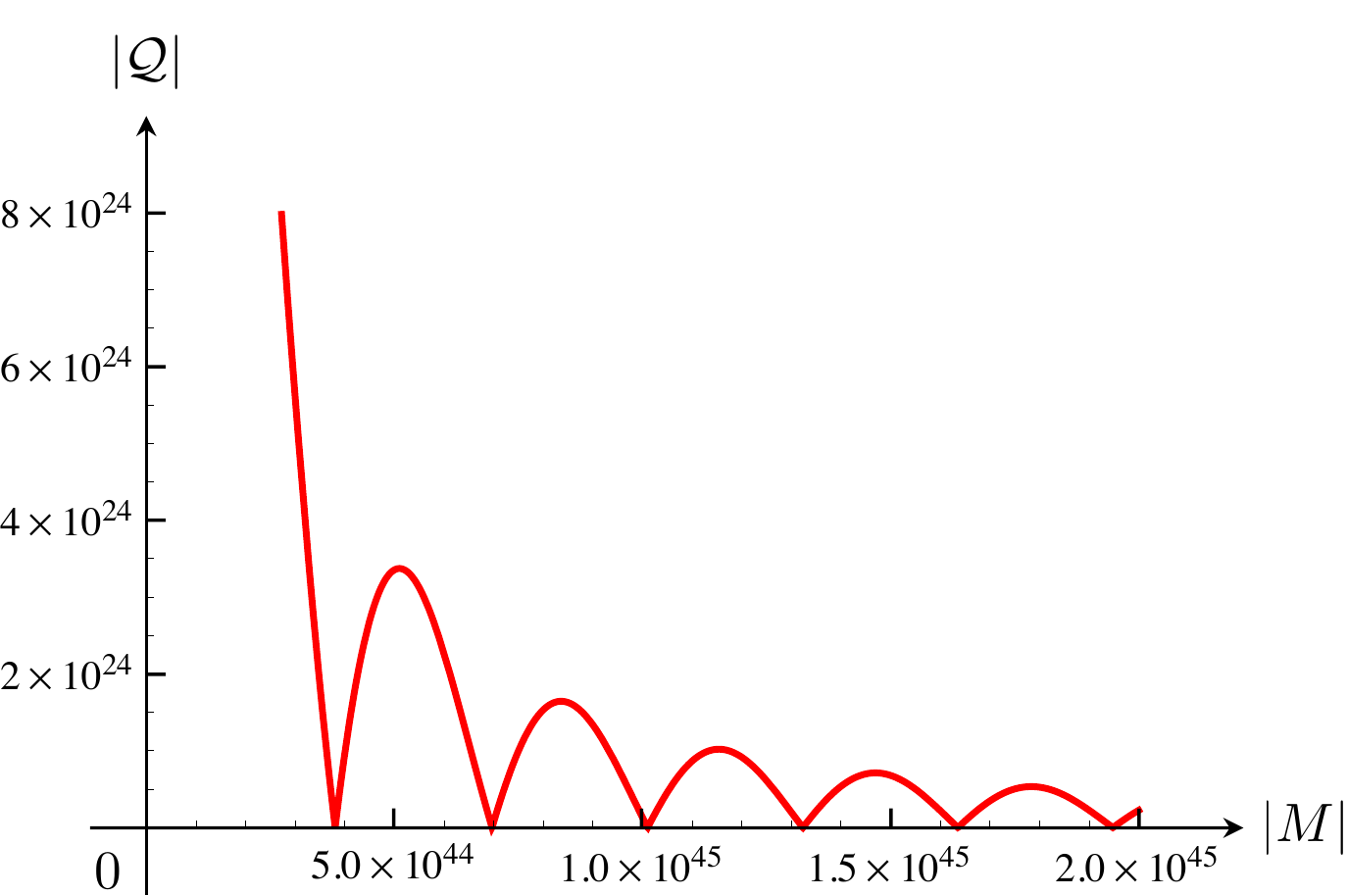}
\end{center}
\caption{The function \eqref{eq:Sawk} for pure imaginary $M$ ($m^2>\frac{9}{4}H^2$).
The zero points appear at almost even intervals.} \label{fig:Seismitoad}
\end{figure}

For pure imaginary $M$ ($m^2>\frac{9}{4}H^2$), this analysis is impossible to carry out with Mathematica due to overflow and underflow problems.
However, since we have
\begin{align}
\left|\frac{P^M_{1/2}\bigr(\cosh(k\xi)\bigr)}{P^M_{1/2}\bigr(\cosh(k(L+\xi))\bigr)}\right|
\lesssim 10^{-8}, \label{eq:Tirtouga}
\end{align}
we can approximately determine the masses $m_n$ by solving
\begin{align}
\frac{Q^M_{1/2}\bigr(\cosh(k\xi)\bigr)}{Q^M_{1/2}\bigr(\cosh(k(L+\xi))\bigr)}=0. \label{eq:Carracosta}
\end{align}
For convenience, we denote the left-hand side of \eqref{eq:Carracosta} by $\Q(M;k\xi,k(L+\xi))$.
Then, we can numerically analyze
\begin{align}
\left|\Q\Bigr(M;k\xi,k(L+\xi)=39+\log\left|\e^{k\xi}-\e^{-k\xi}\right|\Bigr)\right| \label{eq:Sawk}
\end{align}
for $k|\xi|\sim 102$, and the result is shown in Fig.\ \ref{fig:Seismitoad}.\footnote
{Actually, the zeroes of $|\Q|$ in Fig.\ \ref{fig:Seismitoad} are those of the numerator of $|\Q|$.}

\begin{figure}[t]
\begin{center}
\includegraphics[width=10cm,bb=0 0 400 280]{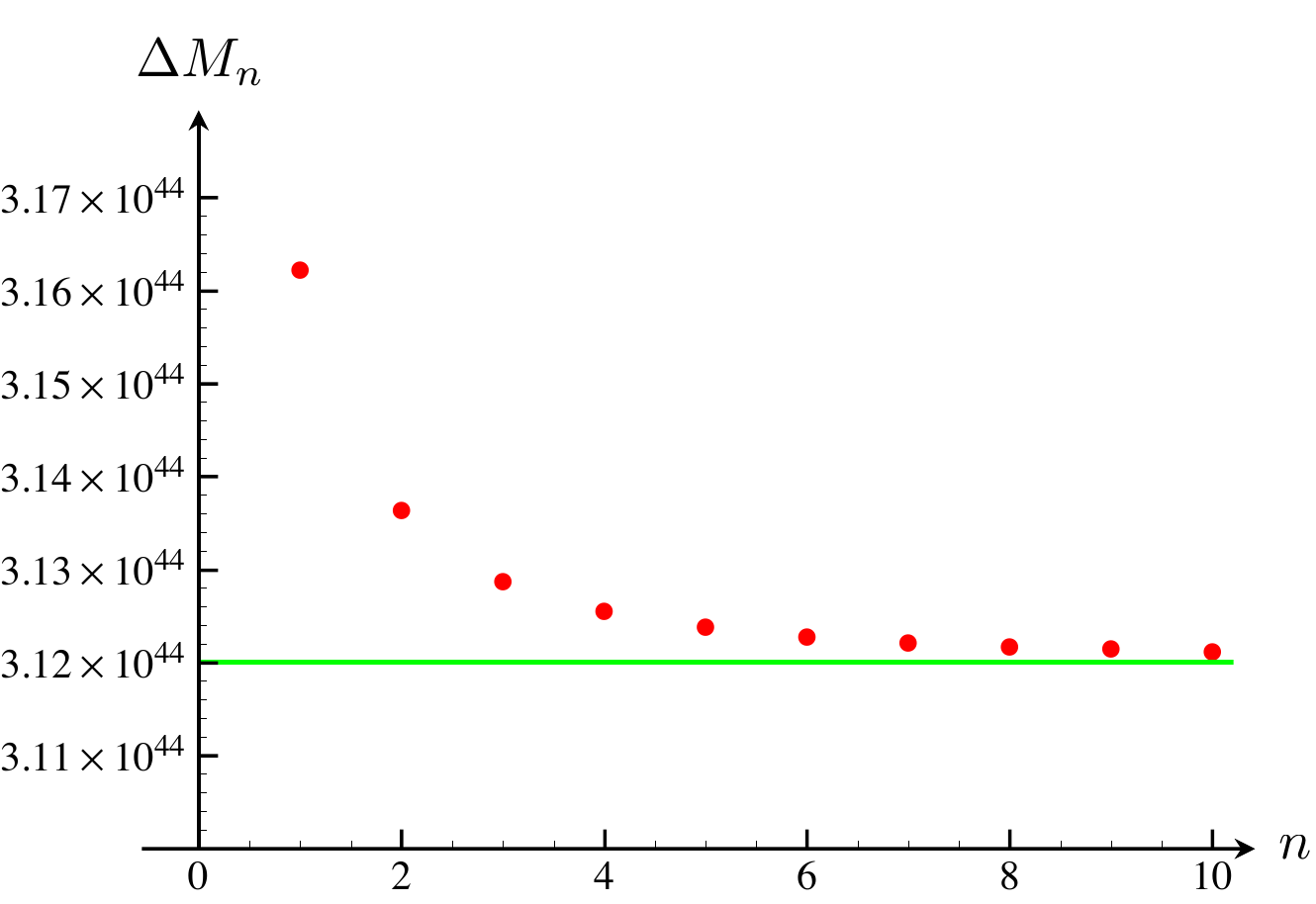}
\end{center}
\caption{The dependence of the intervals $\Delta M_n$ on $n$ for $n=1,2,...,10$ (red dots).
For large $n$, we can realize that $\Delta M_n$ converges to a constant, $(\pi/2)\times \e^{102}=3.12002\times 10^{44}$ (green line).} \label{fig:Gothita}
\end{figure}

From Figs.\ \ref{fig:Palpitoad} and \ref{fig:Seismitoad}, we realize that only the zero-mode, $M_0=3/2$, exists in the range $0\leq m\leq \frac{3}{2}H$, and the zeroes in the range $m>\frac{3}{2}H$ appear at almost even intervals $\Delta M_n=|M_{n+1}|-|M_n|$.
To be exact, $\Delta M_n$ depends on $n$, and for larger $n$, it seems to converge to a constant (see Fig.\ \ref{fig:Gothita}).
The value of $M_1$ corresponding to the mass $m_1$ of the first massive mode and that of $\Delta M_n$ ($n\gg 1$) are determined as\begin{align}
|M_1|=3.80539\times 10^{44}, \qquad
\Delta M_n\simeq 3.12002\times 10^{44}, \label{eq:Conkeldurr}
\end{align}
and accordingly, $m_1$ and the mass intervals $\Delta m_n=m_{n+1}-m_n$ are as
\begin{align}
m_1\sim 3.81\times 10^{2}\,{\rm GeV}, \qquad
\Delta m_n\sim 3.12\times 10^{2}\,{\rm GeV}. \label{eq:Hydreigon}
\end{align}
Here, we have used \eqref{eq:Archeops} and $H_0\sim 10^{-42}\,{\rm GeV}$.
These results are the same as in the RS case (see \eqref{eq:Spearow} of {\S}2.3).

\subsubsection{Tachyonic modes}

From Fig.\ \ref{fig:Palpitoad}, we see that another zero point $M=5/2$ exists in the range $m^2\leq \frac{9}{4}H^2$.
In addition, more zero points exist at $M=7/2,9/2,11/2,...$, though not shown in Fig.\ \ref{fig:Palpitoad}.
The masses $m$ corresponding to these $M$'s are pure imaginary and ``tachyonic''!
If these tachyonic modes really existed, the present model would fail.
Fortunately, these modes do not actually exist.
The origin of the problem is that $P^{(2r+1)/2}_{3/2}(\cosh w)$ is proportional to $Q^{(2r+1)/2}_{3/2}(\cosh w)$ for an integer $r\geq2$, and \eqref{eq:Pidove} is automatically satisfied.
For these exceptional values of $M$, we must prepare two special independent solutions to \eqref{eq:Chandelure}.
For $M_{-1}=5/2$ ($m_{-1}=2iH$), the general solution to \eqref{eq:Chandelure} is given by
\begin{align}
\psi_{-1}(y)=\frac{\sinh^{\frac{5}{2}}(k(|y|+\xi)}{\cosh^5(k(|y|+\xi)}
\Bigr[a_{-1}+b_{-1}\bigr\{12|y|-8\sinh(2|y|)+\sinh(4|y|)\bigr\}\Bigr]. \label{eq:Volcarona}
\end{align}
However, this cannot satisfy the boundary conditions \eqref{eq:Musharna} except for the trivial case $a_{-1}=b_{-1}=0$, implying that the tachyonic mode with $M_{-1}=5/2$ does not exist.
The same is expected to be true for other possible tachyonic modes.


\subsection{Parameter $\xi$}

In this last subsection, we discuss the importance of the parameter $\xi$.
As we saw in {\S}3.1, it came from the Einstein equation as an integration constant.
Such a parameter can also appear in the RS  model, though we did not consider it.
If it is included, the solution $A(y)$ \eqref{eq:Venusaur} is modified as
\begin{align}
A(y)=k|y|+\xi,
\end{align}
and the warp factor $\e^{2A(y)}$ as
\begin{align}
\e^{-2A(y)}=\e^{-2(k|y|+\xi)}=\e^{-2\xi}\cdot \e^{-2k|y|}.
\end{align}
This implies that the parameter $\xi$ causes only a constant multiplication to the warp factor, which can be offset by a rescaling of $x^\mu$.
Therefore, we do not need to consider the parameter $\xi$ in the RS model.

On the other hand, in our model, we cannot offset the parameter $\xi$.
Moreover, for $\Lambda_5<0$, $k\xi$ is related to the Hubble parameter $H$.
Restoring $H$ and $M_{\rm pl}$ in \eqref{eq:Escavalier}, the relation between $H$ and $k\xi$ is given by
\begin{align}
\left(\frac{M_{\rm pl}}{H}\right)^2\sim
\frac{1}{2}\left\{\sinh\Bigr(2\left(k\xi+39+\log\left|1-\e^{-2k\xi}\right|\right)\Bigr)-\sinh(2k\xi)\right\}-39-\log\left|1-\e^{-2k\xi}\right|. \label{eq:Bisharp}
\end{align}
From this relation, we find that another 4-dimensional large ($\sim10^{122}$) hierarchy between the Hubble parameter $H_0$ and the Planck scale $M_{\rm pl}$ is realized by the $\cO(10^2)$ hierarchy, $k|\xi|\sim 102$, between the 5-dimensional quantities $k$ and $1/\xi$.

From the above argument, we see that $k|\xi|$ must not be zero.
This is consistent with the requirement from \eqref{eq:Dewott}; $\xi=0$ implies that the tension of brane 1 located at $y=y_1=0$ becomes infinite.
Hence, the non-zero $\xi$ keeps the model {\it sound}.


\subsubsection{Negative $\xi$ and the zero point of the warp factor $\e^{2A(y)}$}

In the above discussion, there is no way to determine the sign of $\xi$.
In other words, we can carry out the analysis of the Kaluza-Klein modes for the both solutions of \eqref{eq:Pansear}.
For negative $\xi$, the warp factor $\e^{2A(y)}$ has a zero at $y=-\xi$, namely, the 4-dimensional space-time shrinks to a point there.
From the geodesic equation, we can show that particles go through this point.
However, we can also show that the action of the fluctuation $h_{\mu\nu}(x,y)$ is divergent for negative $\xi$.
In fact, using \eqref{eq:Chandelure} for $\psi_n(y)$, we find that the quadratic part of the 4-dimensional fields $\phi^{(n)}_{\mu\nu}(x)$ in \eqref{eq:Simisear} is multiplied by
\begin{align}
\int^{L}_{0} dy\,\e^{-A(y)}\psi^2_n (y).
\label{eq:Cofagrigus}
\end{align}
This integral is divergent at $y=-\xi$, since both $\dis \e^{-A(y)}=1/\left|\sinh(k(|y|+\xi)\right|$ and $\psi_n ^2(y)$ are divergent at $y=-\xi$.
(The mode $\psi_n(y)$ \eqref{eq:Panpour} can be chosen to be real.)
Thus, we conclude that the negative $\xi$ case must be excluded, and, among the two candidates of \eqref{eq:Pansear}, only $(k\xi,kL)\sim (+102,39)$ is the acceptable one.


\subsubsection{Dependences of $m_1$ and $\Delta m_n$ on $k\xi$ (or $H_0$)}

\begin{figure}[t]
\begin{minipage}{0.5\hsize}
\begin{center}
\includegraphics[width=7.7cm,bb=0 0 400 280]{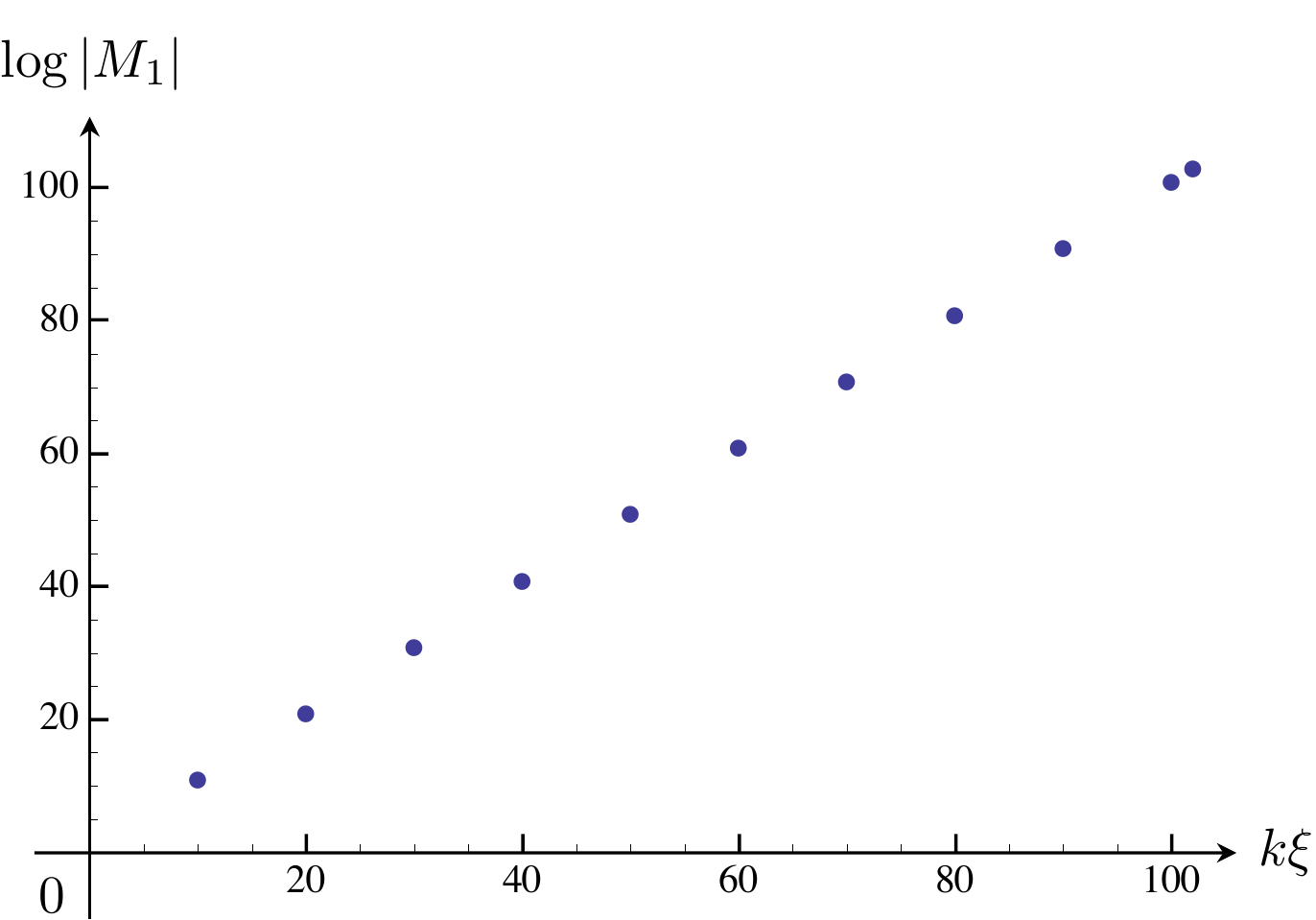}
\end{center}
\end{minipage}
\begin{minipage}{0.5\hsize}
\begin{center}
\includegraphics[width=7.7cm,bb=0 0 400 280]{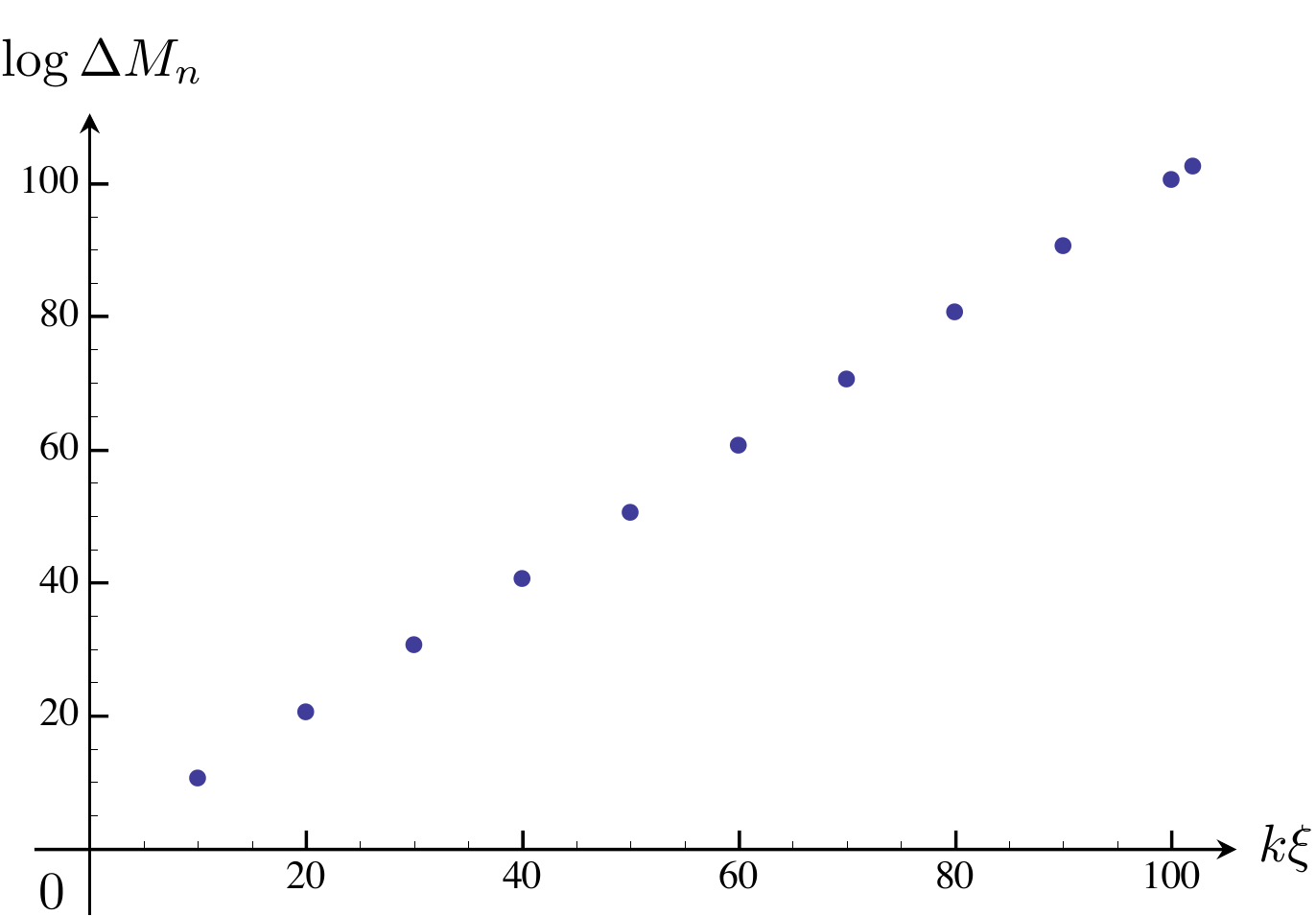}
\end{center}
\end{minipage}
\caption{The plots of $\log|M_1|$ (left figure) and $\log\Delta M_n$ for large $n$ ($n\simeq 200$) (right figure) at $k\xi=10,20,...,100,102$.
We realize that both of $|M_1|$ and $\Delta M_n$ are monotonically (exponentially) increasing functions of $k\xi$.}\label{fig:Axew}
\end{figure}

\begin{figure}[t]
\begin{minipage}{0.5\hsize}
\begin{center}
\includegraphics[width=7.7cm,bb=0 0 400 280]{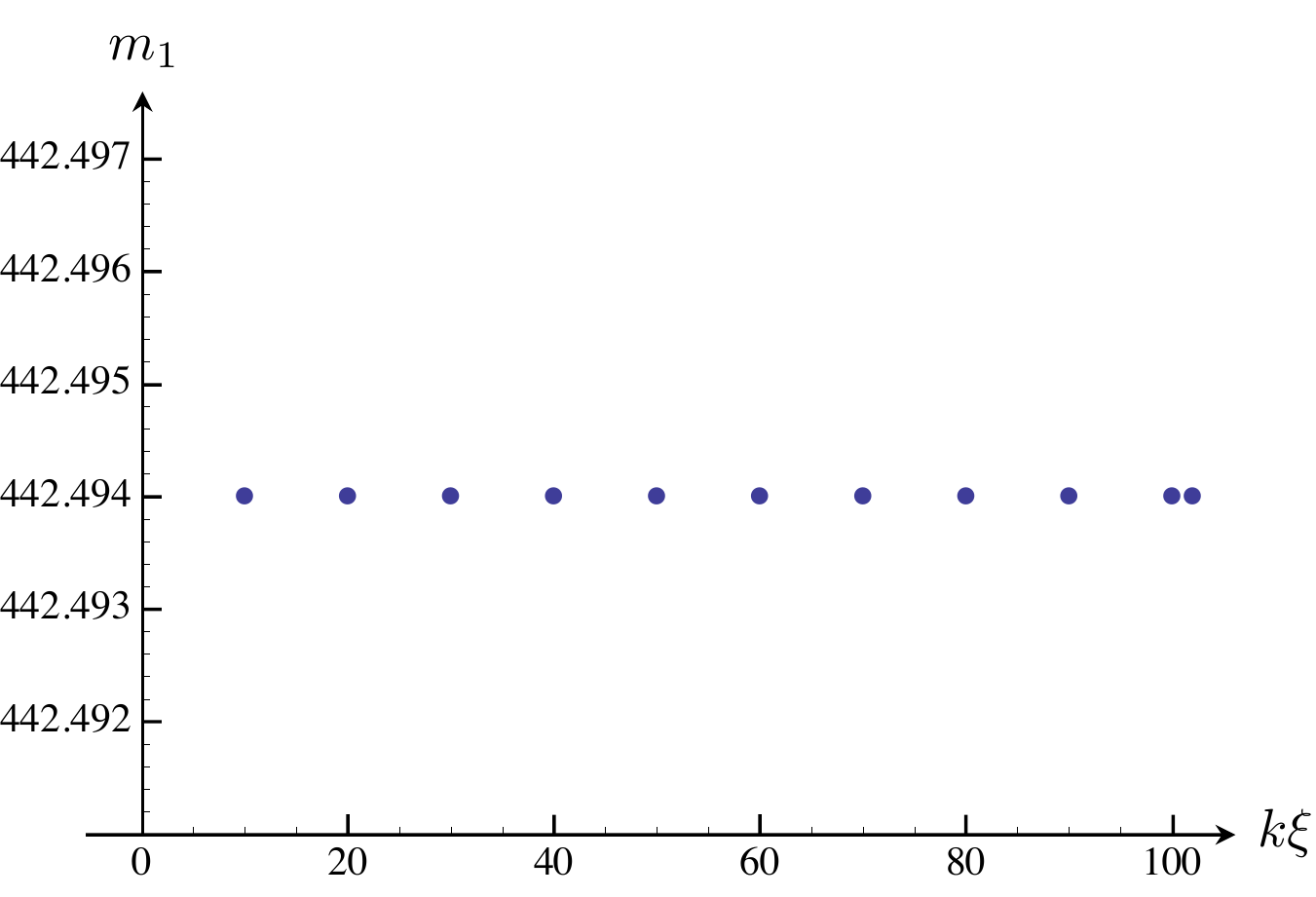}
\end{center}
\end{minipage}
\begin{minipage}{0.5\hsize}
\begin{center}
\includegraphics[width=7.7cm,bb=0 0 400 280]{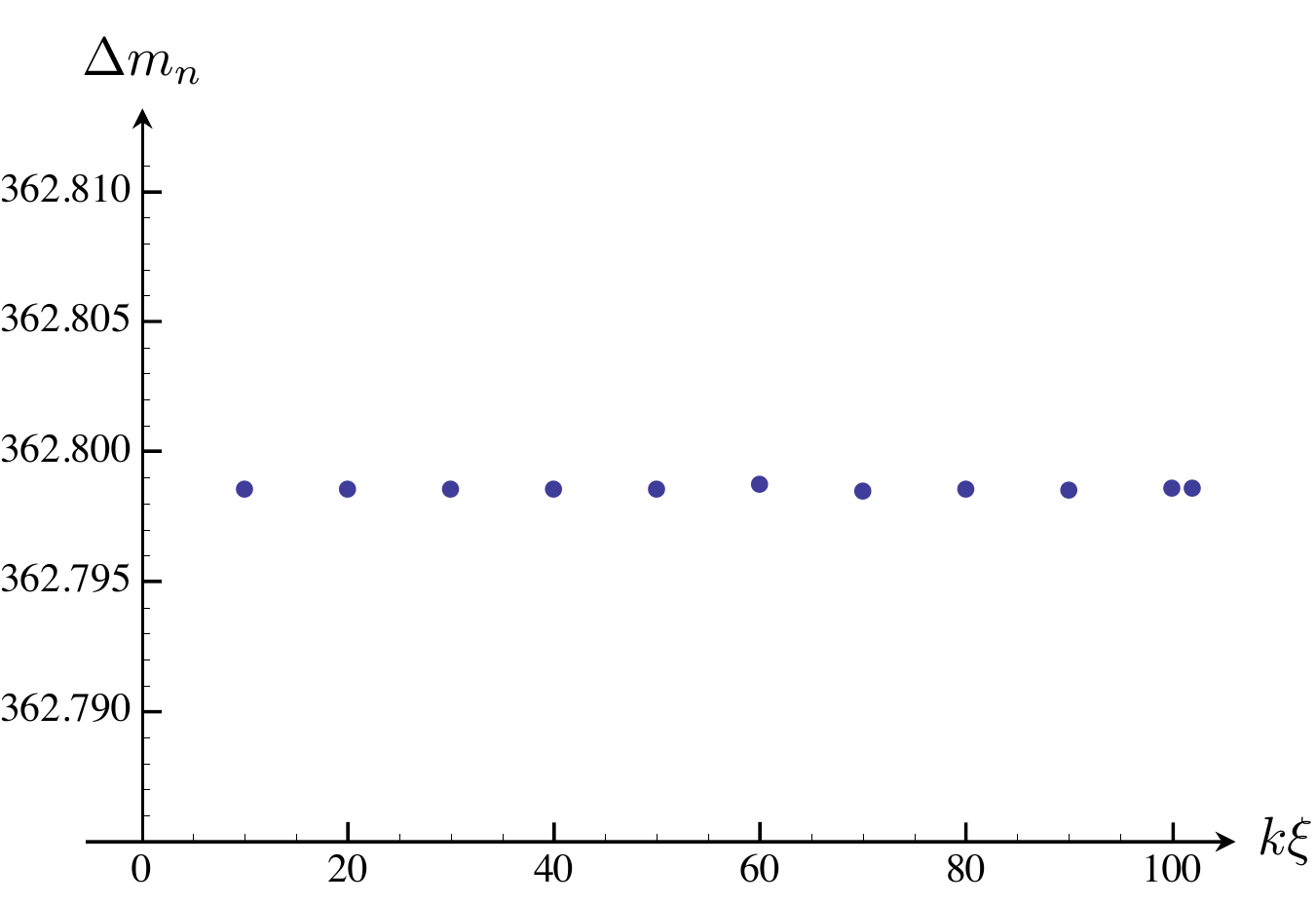}
\end{center}
\end{minipage}
\caption{The plots of $m_1$ (left figure) and $\Delta m_n$ for large $n$ ($n\simeq 200$) (right figure) at $k\xi=10,20,...,100,102$.
We realize that both of $m_1$ and $\Delta m_n$ are almost independent of $k\xi$.
Since we have obtained the values of the Hubble parameter $H_0$ for various $k\xi$ by using \eqref{eq:Bisharp}, the values of the plots are a bit different from \eqref{eq:Hydreigon}.}\label{fig:Fraxure}
\end{figure}

In {\S}3.3.2, we determined the mass $m_1$ of the first massive Kaluza-Klein mode and the intervals of the mass spectrum $\Delta m_n$ for $k\xi \sim 102$, which is determined by the observed value of the Hubble parameter $H_0\sim 10^{-42}\,\rm GeV$.
Here, let us consider how $m_1$ and $\Delta m_n$ depend on $k\xi$ or equivalently on $H_0$.
This is to consider the RS limit of $H_0\rightarrow 0$.
Keeping the 4-dimensional Planck mass $M_{\rm pl}$ a constant, $M_{\rm pl}\sim 10^{19}\,\rm GeV$, the relation between $k\xi$ and $H_0$ is given by \eqref{eq:Bisharp}.
In Fig.\ \ref{fig:Axew} (Fig.\ \ref{fig:Fraxure}), we give the plots of $\log |M_1|$ and $\log \Delta M_n$ ($m_1$ and $\Delta m_n$) as functions of $k\xi$.

\begin{figure}[t]
\begin{minipage}{0.5\hsize}
\begin{center}
\includegraphics[width=7.7cm,bb=0 0 400 280]{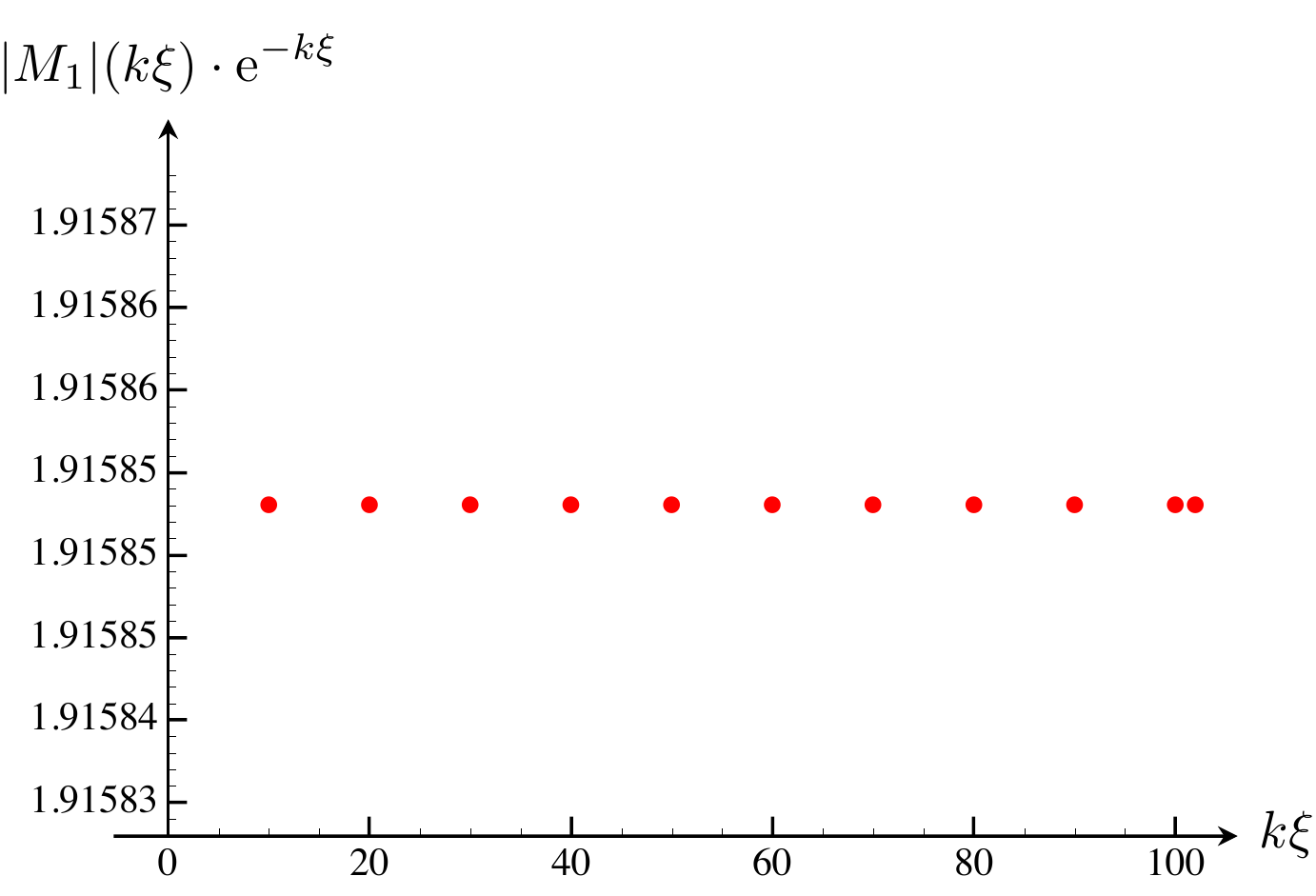}
\end{center}
\end{minipage}
\begin{minipage}{0.5\hsize}
\begin{center}
\includegraphics[width=7.7cm,bb=0 0 400 280]{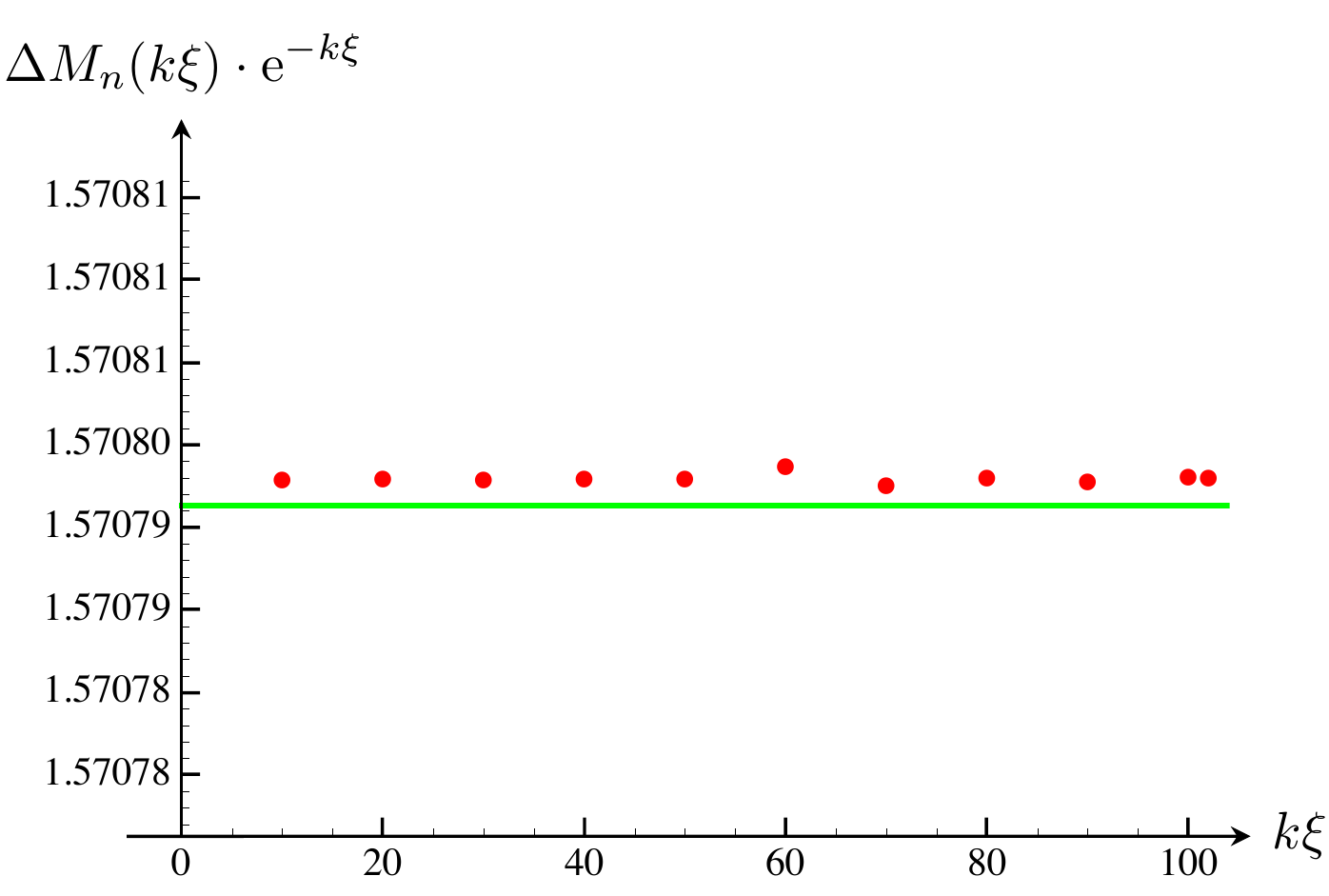}
\end{center}
\end{minipage}
\caption{The plots of $|M_1|(k\xi)\cdot \e^{-k\xi}$ (left figure) and $\Delta M_n(k\xi)\cdot \e^{-k\xi}$ (right figure) for large $n$ ($n\simeq 200$) and $k\xi=10,20,...,100,102$.
The green line shows the value $\pi/2$.}\label{fig:Haxorus}
\end{figure}

From Fig.\ \ref{fig:Axew}, we realize that $|M_1|$ and $\Delta M_n$ are both monotonically increasing functions of $k\xi$.
However, from Fig.\ \ref{fig:Fraxure}, we also see that $m_1$ and $\Delta m_n$ are both almost independent of $k\xi$.
This means that the product of $H_0$ and $|M_1|$ and that of  $H_0$ and $\Delta M_n$ are constants independent of $k\xi$:
\begin{align}
m_1\simeq H_0(k\xi)\times |M_1|(k\xi)= \text{const},\qquad
\Delta m_n\simeq H_0(k\xi)\times \Delta M_n(k\xi)= \text{const}, \label{eq:Emolga}
\end{align}
where we have made the large $m_n$ approximation of \eqref{eq:Archeops}, and have written explicitly that $H_0$, $|M_1|$, and $\Delta M_n$ depend on $k\xi$.
For large $k\xi$ ($\e^{k\xi}\gg1$), the relation \eqref{eq:Bisharp} between $k\xi$ and $H_0(k\xi)$ is approximately expressed as
\begin{align}
M_{\rm pl}^2\sim \frac{\e^{78}}{4}\times H_0^2(k\xi)\cdot \e^{2k\xi}
\simeq \e^{78}k^2. \label{eq:Gothitelle}
\end{align}
Here, we have used the second equation of \eqref{eq:Virizion}.
Now, we keep $M_{\rm pl}$ a constant, and then, from \eqref{eq:Emolga} together with \eqref{eq:Gothitelle}, we obtain
\begin{align}
|M_1|(k\xi)\propto \Delta M_n(k\xi)\propto \e^{k\xi}. \label{eq:Cinccino}
\end{align}
Fig.\ \ref{fig:Haxorus} shows $|M_1|(k\xi)\cdot \e^{-k\xi}$ (left figure) and $\Delta M_n(k\xi)\cdot \e^{-k\xi}$ for large $n$ ($n\simeq 200$) (right figure) at $k\xi=10,20,...,100,102$.
From this figure, we reconfirm that $|M_1|(k\xi)$ and $\Delta M_n(k\xi)$ are almost proportional to $\e^{k\xi}$, and moreover, we find that $\Delta M_n(k\xi)\cdot \e^{-k\xi}$ is almost equal to $\pi/2$.
Hence, for pure imaginary $M$ and large $k\xi$, we expect that the function $\B(M;k\xi,k(L+\xi))$ can be expressed as
\begin{align}
\B(M;k\xi,k(L+\xi))\propto \sin\left(2|M|\cdot \e^{-k\xi}\right)\quad (M\neq 0).
\end{align}

\subsubsection{RS limit}

Finally, we consider a limit where we can obtain the results of the RS model from those of our model.
(This limit is often called ``the RS limit''.)
Here, we consider the situation where the warp factor is normalized at the TeV brane position, and all the 5-dimensional quantities are almost of the same order $10^2\,\rm GeV$.
Taking the limit of $H_0\rightarrow +0$ in the metric ansatz \eqref{eq:Victini} of our model, we obtain the metric \eqref{eq:Ivysaur} of the RS model except for the sign of $A(y)$.
In taking this limit, we must fix $H_0\e^{k\xi}$ a constant as seen from \eqref{eq:Gothitelle}, namely, we must take the limit $\xi \rightarrow \infty$.
Then, let us check whether this limit, $H_0\rightarrow +0$, $\xi\rightarrow \infty$ with fixed $H_0\e^{k\xi}=2k$, is just the RS limit.
Applying this limit to the warp factor of \eqref{eq:Terrakion}, we find that
\begin{align}
\left(\frac{\sinh(k(|y|+\xi))}{\sinh(k\xi)}\right)^2
=\left(\e^{k|y|}\cdot \frac{1-\e^{-2k(|y|+\xi)}}{1-\e^{-2k\xi}}\right)^2
\,\overset{\xi\rightarrow \infty}{\longrightarrow}\,\e^{2k|y|}, \label{eq:Keldeo}
\end{align}
which is equal to the warp factor of the RS model (see \eqref{eq:Ivysaur} and \eqref{eq:Venusaur}).\footnote
{As stated in footnote \ref{fn:Chansey} on page $5$, reversing the sign of $k$ is essentially equivalent to replacing $y$ with $L-y$, which means exchanging the positions of the Planck brane and the TeV brane.
Actually, in the RS model, we located the Planck brane (the TeV brane) at $y=0$ ($y=L$).
On the contrary, in our model, the positions of the two branes are exchanged.
To exchange the brane positions in our model, we should do the same replacement of $y\rightarrow L-y$.
Then, \eqref{eq:Keldeo} is modified to
\begin{align*}
\left(\frac{\sinh(k(-|y|+L+\xi))}{\sinh(k\xi)}\right)^2
\,\overset{\xi\rightarrow \infty}{\longrightarrow}\,\e^{2k(L-|y|)},
\end{align*}
which is just equal to \eqref{eq:Dragonite}.}


\setcounter{section}{3}
\setcounter{subsection}{0}
\setcounter{figure}{0}

\section{Possibility of making the Hubble parameter time-dependent}

In this paper, we focused for simplicity only on our current universe with the observed Hubble parameter, and did not consider the complicated time evolution of the universe.
A possible way to make the Hubble parameter $H$ time-dependent would be to promote the brane interval $L$, which is a constant in the present model, to a dynamical variable.
This also makes the weak scale, namely, the scale $v_1$ appearing in \eqref{eq:Patrat}, time-dependent.
Furthermore, to stabilize the model, we must introduce a scalar field (radion) in the bulk\cite{Goldberger:1999uk,Goldberger:1999un}, and the analyses of the Einstein equation, the hierarchy problem, the Kaluza-Klein modes, and so on, will become more complicated.
(Then, the model must be called ``the thick brane model''.)

On the other hand, one might think that another way to make the Hubble parameter $H$ time-dependent is to allow the warp factor to depend both on $t$ and $y$.
Under the metric assumption,
\begin{align}
ds^2= \e^{2B(y,t)}\left\{-dt^2+a^2(t)\eta_{ij}dx^idx^j\right\}+dy^2, \label{eq:Celebi}
\end{align}
the Einstein equation leads to
\begin{align}
B(y,t)=A(y)+\omega(t),\qquad
\omega(t)+\log a(t)=H\int^t dt'\,\e^{\omega(t')}, \label{eq:Jirachi}
\end{align}
where $H$ is a constant, $A(y)$ is given by \eqref{eq:Tepig}, and $\omega(t)$ is an arbitrary function of $t$.
Then, let us introduce a new coordinate $\tau$ defined by
\begin{align}
\tau=\int^t dt'\,\e^{\omega(t')}, \label{eq:Deoxys}
\end{align}
and realize $d\tau^2=\e^{2\omega(t)}dt^2$.
In this manner, the metric \eqref{eq:Celebi} becomes
\begin{align}
ds^2=\e^{2A(y)}\left\{-d\tau^2+\e^{2H\tau}\eta_{ij}dx^idx^j\right\}+dy^2,
\end{align}
which is equivalent to the metric of our model with a constant Hubble parameter.
Hence, we realize that it is meaningless to make the warp factor time-dependent.


\section*{Acknowledgements}
I would like to thank H. Hata, Y. Hosotani, K. Oda, S. Yamaguchi, and the other members of Osaka Univ. Particle Physics Theory group for valuable discussions and useful comments.



\end{document}